\documentclass{article}
\usepackage{spconf,amsmath,graphicx,color}
\usepackage{booktabs}
\title{Multi-Speaker Expressive Speech Synthesis via Multiple factors Decoupling}
\name{Xinfa Zhu, Yi Lei, Kun Song, Yongmao Zhang, Tao Li, Lei Xie$^{*}$\thanks{* Corresponding author. This work was supported by National Key R\&D Program of China, under Grant No. 2020AAA0108600.}}
\address{
  Audio, Speech and Language Processing Group (ASLP@NPU)\\School of Computer Science,
  Northwestern Polytechnical University, Xi’an, China}
\begin{document}
\ninept
\maketitle

\begin{abstract}
\vspace{-5pt}
This paper aims to synthesize the target speaker's speech with desired speaking style and emotion by transferring the style and emotion from reference speech recorded by other speakers.
We address this challenging problem with a two-stage framework composed of a text-to-style-and-emotion (Text2SE) module and a style-and-emotion-to-wave (SE2Wave) module, bridging by neural bottleneck (BN) features. 
To further solve the multi-factor (speaker timbre, speaking style and emotion) decoupling problem, we adopt the multi-label binary vector (MBV) and mutual information (MI) minimization to respectively discretize the extracted embeddings and disentangle these highly entangled factors in both Text2SE and SE2Wave modules. Moreover, we introduce a semi-supervised training strategy to leverage data from multiple speakers, including emotion-labeled data, style-labeled data, and unlabeled data. 
To better transfer the fine-grained expression from references to the target speaker in non-parallel transfer, we introduce a reference-candidate pool and propose an attention-based reference selection approach. Extensive experiments demonstrate the good design of our model.
\vspace{-3pt}

\end{abstract}
%\renewcommand{\thefootnote}{\fnsymbol{footnote}}
%\footnotetext{* Corresponding author.}
%
\vspace{-10pt}
\section{Introduction}
\vspace{-10pt}
\label{sec:intro}
In recent years, neural text-to-speech (TTS) synthesis has made rapid progress regarding quality and naturalness \cite{Tan2021ASO,Ren2021FastSpeech2F,DBLP:conf/nips/KongKB20,Kim2021ConditionalVA}.
With the wide applications of TTS, there have been increasing demands for robust expressive speech synthesis systems to provide more human-like speech in diverse scenarios. 
In previous works of expressive speech synthesis, speech expressiveness usually refers to specific speaking styles or emotional expressions associated with speech~\cite{Li2022StyleTTSAS, Sorin2020, Lei2022MsEmoTTSME, Liu2022RefereeTR}.
%Nevertheless, there is still a vast gap between synthetic and natural human speech, especially speech with expressive emotions and styles. The synthesized speech is often in a fixed style, and the expressed emotion is not delicate enough. We define scenes, such as poetry and fairy tales, as styles and describe the attitudes and intentions, such as anger and sadness, as emotions. Generating highly expressive speech with emotions and styles is vital to improving the human-computer interaction experience. 

A straightforward approach~\cite{Lee2017EmotionalEN,Li2018EmphaticSG,zhang2019learning,DBLP:conf/iscslp/LiYXX21} to synthesize expressive speech for a specific speaker is to train a TTS system with his/her expressive training speech. However, it can not be generalized to target speakers without expressive training data, which is hard to obtain for each target speaker. Therefore, transferring emotion or style from a source speaker to target speakers is a feasible strategy, where the source speaker has expressive speech while the target speaker only has neutral speech. The key factor for emotion or style transfer is to decouple the speaker timbre and expressive aspects from speech~\cite{Wang2021AdversariallyLD,Bian2019MultireferenceTB,Liu2022RefereeTR,9874835}. Speaker timbre essentially reveals the physiological characteristics of individual's vocal tract, while emotion and speaking style are more behavioral. However, it is not a trivial task as these aspects are highly entangled in the speech signal.
Some works~\cite{Sorin2020,Ma2019NeuralTS,Li2022CrossSpeakerED} try to disentangle speaker timbre and style or emotion in the latent space to conduct style or emotion transfer. However, decoupling approaches in latent space usually need to carefully select an appropriate reference signal during inference. These reference-based style transfer methods always face a trade-off between expressiveness and speaker similarity, which leads to a vast performance gap between synthetic and natural human speech. 

To solve this problem, some articles adopt the neural network bottleneck (BN) features or Phonetic PosteriorGrams (PPGs) as intermediate representations for decoupling. BN features are a set of activation of nodes over time from a neural network bottleneck layer, while PPGs are obtained by stacking a sequence of phonetic posterior probabilistic vectors from the neural network. BN features and PPGs, usually obtained from a well-trained acoustic model in an automatic speech recognition system, are believed to be linguistic-rich~\cite{acvc,9746709}, speaker-independent~\cite{againvc}, noise-robust~\cite{li2021ppg}, and contain stylistic information such as duration and accent~\cite{huang2021aispeech}. Leveraging the intermediate representations, the style transfer TTS problem can be simplified to a two-stage process, where the first stage mainly manages the style learning from the source speaker while the second stage aims at the target speaker timbre modeling. Through such a two-stage framework, style or emotion transfer can be conducted without a reference signal during inference. Referee~\cite{Liu2022RefereeTR} is a representative work in this direction that adopts PPGs as the intermediate representations connecting the two-stage models for cross-speaker style transfer. 
%while prosody bottleneck~\cite{pan2021cross} shares the similar idea in style and speaker timbre disentanglement.

%The explicit prosody features are proposed~\cite{pan2021cross} to disentangle the speaker and style in cross-speaker style transfer TTS. Referee ~\cite{Liu2022RefereeTR} is proposed to adopt Phonetic PosteriorGrams (PPGs) as the intermediate features, extracted from Automatic Speech Recognition (ASR) systems and speaker-independent. Referee consists of two cascading models for providing style-related representations and waveforms of target speakers, respectively. 

The above style transfer approaches usually have an unclear definition of style and emotion and sometimes consider emotion as a type of speaking style. Whereas, this indiscriminate treatment of style and emotion restricts them to be extended to diverse scenarios requiring both multiple emotions and styles, which is common in real applications. Speaking style is a general distinctive style of speech in different usage scenarios, such as news reading, storytelling, poetry recitation, and conversation. Even for storytelling, telling different stories (such as fairy tales and novels) may use different speaking styles. By contrast, emotion mainly reflects the mood state of the speaker, related to attitudes and intentions, conveyed differently in each utterance, such as happy, angry, sad, etc. Moreover, different emotions can be expected in different places in an audio stream with a global speaking style (such as storytelling).

%the emotion transfer to target speakers, who requires different styles to adapt to diverse scenarios.

%Whereas, in the real-applications, speakers may have their own expression way to 
%tradition-->refree-->bian 

%Firstly, we would like to define the multiple factors to be decoupled as follows: 
%\begin{itemize}
%    \item \textbf{Speaking style.} The speaking style in this paper stands for how the speech is expressed and related to the speaking situation and overall scene, such as poetry, fairy tales, chat, etc.
%    \item \textbf{Emotion.} The emotion here describes the mood state, related to attitudes and intentions, conveyed in each utterance, such as happy, angry, sad, etc.
%    \item \textbf{Speaker timbre.} The speaker timbre is the vocal print, which is the identity of each speaker.
%\end{itemize}
%To address this problem and realize both speaking style and emotion transfer in expressive speech synthesis, this paper propose a multi-factor decoupling approach based on bottleneck features.
%Given the above description, in this paper, we try to conduct the multi-speaker expressive speech synthesis via decoupling multiple factors, i.e. style, emotion, and speaker from speech. 

In this paper, we focus on both speaking style and emotion transfer in multi-speaker expressive speech synthesis. The challenges for building such a multi-factor system are threefold. First, explicitly decoupling multiple factors -- speaking style, emotion, and speaker timbre -- is more difficult as style and emotion patterns are both reflected in speech prosody and thus highly entangled. Second, it is also difficult to obtain expressive data labeled with both emotion and speaking style. %Third, the approach on expressive TTS with multi-factor delivery needs to be elaborately designed because the aforementioned reference-based style transfer method faces the dilemma on expressiveness and speaker similarity while reference-free style tag based approach ignores fine-grained prosody modeling with over-averaged style or emotion delivery in synthetic speech.  
Third, the reference-based model mentioned above has a mismatch problem in practical non-parallel transfer scenarios, i.e. the novel text content during inference is different from the reference signal selected in the training data, leading to performance degradation, which is severe in multi-speaker expressive speech synthesis~\cite{Bian2019MultireferenceTB} and may become more critical in our multi-factor case.

Our proposed approach leverages the advances of the two-stage framework with a text-to-style-and-emotion module (\textit{Text2SE}) and a style-and-emotion-to-wave (\textit{SE2Wave}) module. The former predicts linguistic, style, and emotional information embedded in the neural bottleneck (BN) feature, while the latter takes the BN feature as input and produces the target speaker's voice with both stylistic and emotional factors. 
% To address the multi-factor decoupling problem, we explicitly adopt an emotion extractor and a style extractor in the Text2SE module, where multi-label binary vector~(MBV)~\cite{Liu2019UnsupervisedEL} for information discretization is particularly adopted in both extractors and  style and emotion are decoupled by mutual information (MI) minimization~\cite{Cheng2020CLUBAC} across the two extractors. As for the data sparsity problem, we introduce a semi-supervised training strategy to leverage expressive data from multiple speakers, including emotion-labeled data, style-labeled data as well as unlabeled data. Finally for the design of the SE2Wave module, we mange to 
%To tackle these issues, we propose a novel multi-factor decoupling scheme for flexibly controlling and recomposing \textit{style}, \textit{emotion}, and \textit{speaker} in expressive speech synthesis. To achieve a robust decoupling performance, we build a two-stage framework taking the BN, which are speaker-independent bottleneck features, as the intermediate representations connecting two models, namely \textit{Text2SE} and \textit{SE2Wave} model. The \textit{Text2SE} model is to generate the speaker-independent representations (BN) in desired style and emotion. With BN, emotion, and speaker identity as input, the \textit{SE2Wave} model can produce stylistic and emotional speech waveform in the voice of target speakers. 

Based on the two-stage framework, this paper proposes the following designs. To address the multi-factor decoupling problem, we adopt the multi-label binary vector~(MBV)~\cite{Liu2019UnsupervisedEL} and mutual information (MI) minimization~\cite{Cheng2020CLUBAC} to respectively discretize the extracted embeddings and decouple the style, emotion, and speaker factors in the design of both Text2SE and SE2Wave modules. As for the data sparsity problem, we introduce a semi-supervised training strategy to leverage expressive data from multiple speakers, including emotion-labeled data, style-labeled data, and unlabeled data.
%To achieve a meticulous expression of style and emotion with high speaker similarity, we combine the advantages of reference-based and reference-free transfer methods and introduce different embedding extractors in training and fine-tuning procedure. With the different embedding extractors, the system can generate a better prosody and avoid the trivia of manually selecting references for arbitrary input text.
%relieve the problems of both reference-based and reference-free approaches, the trade-off between expressiveness and speaker similarity in reference-based approaches and the over-averaged expressions of reference-free approaches, we introduce different embedding extractors in training and fine-tuning procedure for better prosody generation and avoiding the trivia of manually selecting references.
%Combining the advantages of reference-based and reference-free transfer methods, we introduce different embedding extractors in training and fine-tuning procedure for better prosody generation and avoiding the trivia of manually selecting references.
To eliminate the mismatch problem in non-parallel transfer scenarios, we introduce a reference-candidate pool and propose an attention-based reference selection approach, which reserves the fine-grained prosody from the reference signal and avoids the difficulty of manual reference selection. Extensive experiments demonstrate the good design of our model. We suggest the readers listen to our online demos~\footnote{Demo: https://zxf-icpc.github.io/multi-factor-decoupling/}.

\vspace{-10pt}
\section{PROPOSED APPROACH}
\vspace{-8pt}

\label{sec:pagestyle}

\begin{figure}[htb]
\vspace{-8pt}

\begin{minipage}[b]{1.0\linewidth}
  \centering
  \centerline{\includegraphics[width=\textwidth]{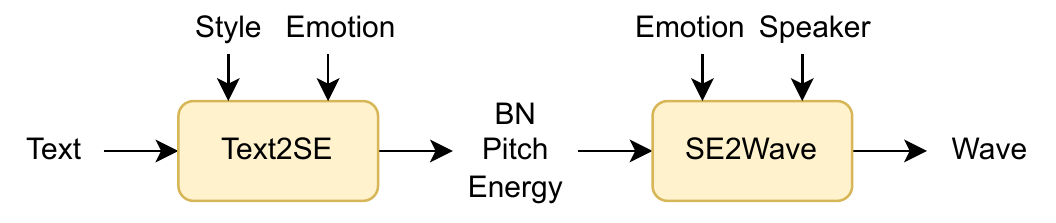}}
%  \vspace{2.0cm}
\end{minipage}
\vspace{-20pt}

\caption{The architecture of proposed approach.}
\label{fig_1}
\vspace{-8pt}

\end{figure}

The proposed approach consists of a Text2SE module and a SE2Wave module, as shown in Figure~\ref{fig_1}. The Text2SE module is to predict BN features, pitch, and energy, which are speaker-independent intermediate representations with style and emotion. The SE2Wave module aims at waveform generation of the target speaker in the desired emotion and style. Note that emotional information is superposed in the procedure of wave generation as the supplement of the fine-grained emotion variations for natural expression delivery. As detailed in Section 2.4, the whole system will go through a 
training stage and a fine-tuning stage, where different embedding extractors are used in the two stages respectively to ensure good performance.
%For the reference-based style/emotion transfer, the style/emotion embedding is extracted from the target mel-spectrogram, which is parallel with text content, at training time. While during inference, the reference signal is usually non-parallel, resulting in a mismatch problem of the extracted embedding between the training and inference process. To relieve the performance degradation brought by this mismatch problem, we propose different embedding extractors for the training and fine-tuning process. 

\vspace{-10pt}
\subsection{The Text2SE module}
\vspace{-5pt}

\begin{figure}[t]
% \vspace{-2pt}

\begin{minipage}[b]{1.0\linewidth}
  \centering
  \centerline{\includegraphics[width=0.8\textwidth,height=0.76\textwidth]{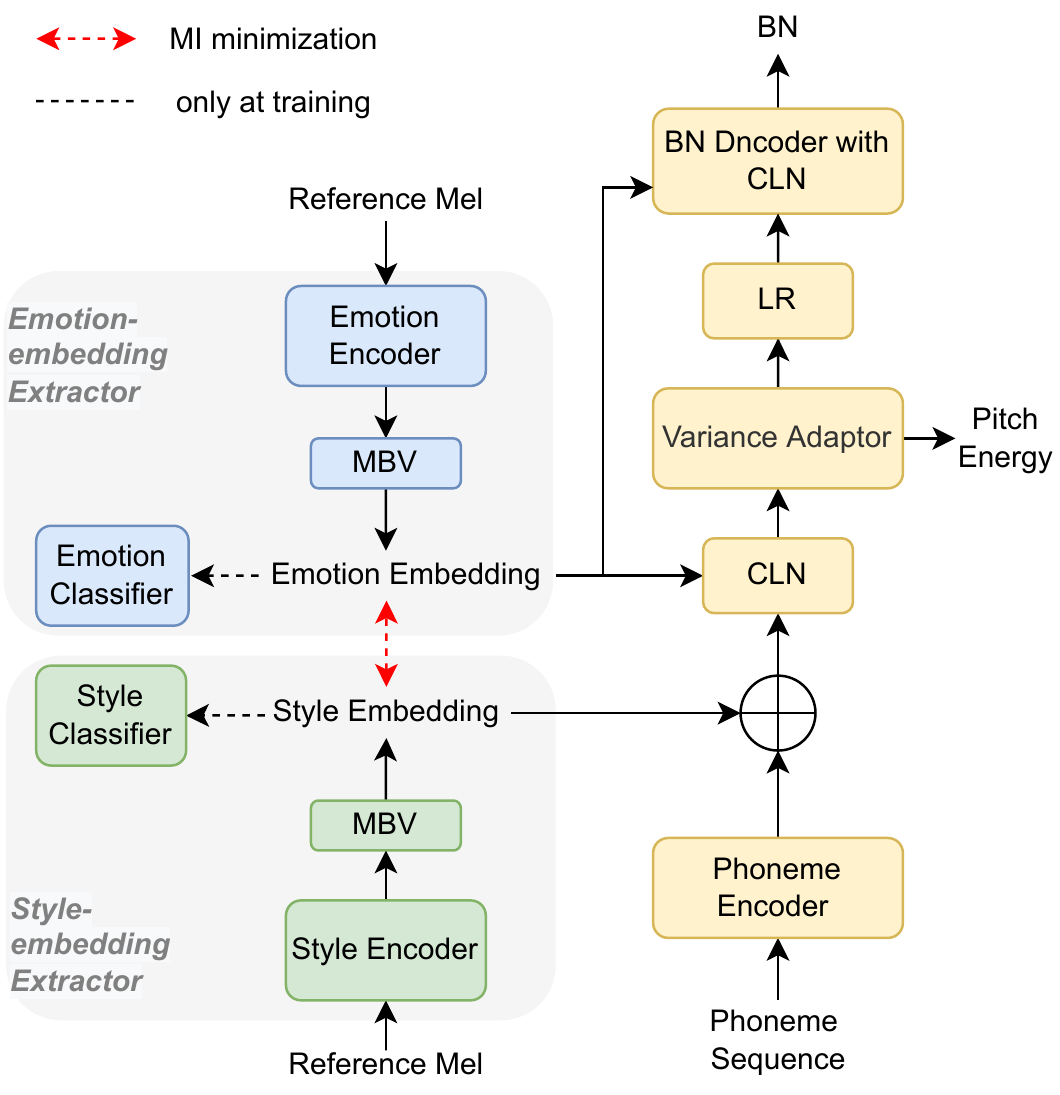}}
\end{minipage}
\vspace{-20pt}
\caption{The Text2SE module architecture.}
\label{fig_2}
\vspace{-20pt}

\end{figure}

As illustrated in Figure~\ref{fig_2}, the Text2SE module is shaped with a backbone model and two embedding extractors for style and emotion respectively. The backbone model is mainly composed of a phoneme encoder, a variance adaptor, and a BN decoder. The goal of this module is to produce the speaker-independent BN features conditioning on the style and emotion embeddings. %For the style and emotion embedding extraction, we utilize multi-speaker training data with style label only, emotion label only, or none of both.

%-----------------------------
During training, the style/emotion embedding extractor contains a style/emotion encoder, a Multi-label Binary Vector~(MBV)~\cite{Liu2019UnsupervisedEL}, and a classifier.
The style/emotion encoder takes the mel-spectrogram as input and then uses an MBV to discretize the output for compression. As a bottleneck layer, MBV with Gumbel-Softmax can reduce the difficulty of multi-factor decoupling and stabilize the adversarial training. To train the embedding extractors, we add classification constraints to the extracted embeddings. Besides, we adopt the variational contrastive log-ratio upper bound (vCLUB)~\cite{Cheng2020CLUBAC} to measure the mutual information between the extracted style and emotion embeddings for sufficiently decoupling style and emotion by mutual information (MI) minimization.

%The extracted emotion embedding is fed into the model through Conditional LayerNorm (CLN) and concatenated with the style embedding and phoneme encoder output for modeling the diverse styles and emotions. The pitch and energy, normalized in the utterance level for eliminating the speaker-related features, are predicted by the variance adaptor. Then the target speaker-independent PPGs with style and emotion are produced by the PPG decoder. The network of each module in the Text2SE follows the structure used in FastSpeech 2~\cite{Ren2021FastSpeech2F}.

The extracted emotion embedding is fed into the backbone model through Conditional LayerNorm (CLN), and the style embedding is first concatenated with the phoneme encoder output and then goes through the CLN. The variance adaptor predicts the pitch and energy normalized in the utterance level for eliminating the speaker-related attributes. Finally, the BN decoder produces the speaker-independent BN with style and emotional information. The backbone network follows the structure used in FastSpeech2~\cite{Ren2021FastSpeech2F}.

%Emotion representation is embedded into the network through Conditional LayerNorm (CLN), and style representation is embedded into the network through contact with the output of the text encoder. We adopt the reference encoder to encode the information of the mel-spectrogram to model diversity of style and emotion. Moreover, The Multi-label Binary Vector(MBV) \cite{Liu2019UnsupervisedEL} compresses the global information obtained by the reference encoder. MBV can not only effectively squeeze the amount of information and reduce the difficulty of decoupling but also stabilize the adversarial training of the model. Besides, there is almost no over-compression because the kind of information it expresses is exponential.

%Typically, data either has a style tag, an emotional tag, or none. The traditional approach that gives unlabeled data a default label ignores the commonalities between unlabeled data and labelled data. For example, some unlabelled data have emotional interpretation during recording, and it is inappropriate to consider it neutral. Therefore, this paper adopts semi-supervised training. The classifier loss of unlabelled data is not calculated, and the model will softly determine what style or emotions it contains.

%Both style representation and emotional representation are extracted from the mel-spectrogram. Although they are constrained by classification, they easily contain each other's information. This paper uses variational contrastive log-ratio upper bound (vCLUB) \cite{Cheng2020CLUBAC}  to measure the mutual information in the extracted style representation and emotion representation and realize the decoupling of style and emotion by mutual information minimization.

The training objective of the Text2SE module $\mathcal{L}_{\mathrm{t2se}}$ is
{
\setlength\abovedisplayskip{2.0pt}
\setlength\belowdisplayskip{2.0pt}
\begin{equation}
    \mathcal{L}_{\mathrm{t2se}}=  \mathcal{L}_{\mathrm{BN}}+ 0.1 \cdot \mathcal{L}_{\text {prosody}} +0.1 \cdot \mathcal{L}_{\mathrm{MI}} 
    + \mathcal{L}_{\mathrm{emo}} + \mathcal{L}_{\mathrm{sty}},
    % \mathcal{L}_{\mathrm{t2se}}= \mathcal{L}_{\mathrm{BN}}+\lambda_{\text {prosody }} * \mathcal{L}_{\text {prosody}} 
    % +\lambda_{\mathrm{MI}} * \mathcal{L}_{\mathrm{MI}} +\lambda_{\mathrm{cls}} * \mathcal{L}_{\mathrm{cls}}
\label{eq:eq1}
\end{equation}
}
where $\mathcal{L}_{\mathrm{BN}}$ is the reconstruction loss of BN, $\mathcal{L}_{\text {prosody}}$ is the loss for predicting pitch and energy, $\mathcal{L}_{\mathrm{MI}}$ is the MI loss between emotion embedding and style embedding, $\mathcal{L}_{\mathrm{emo}}$ and $\mathcal{L}_{\mathrm{sty}}$ are the classification loss of emotion and style embedding.

\vspace{-10pt}
\subsection{The SE2Wave module}
\vspace{-5pt}

Likewise, the SE2Wave architecture is composed of a backbone model and two embedding extractors for speaker and emotion respectively. The backbone is based on VITS~\cite{Kim2021ConditionalVA}, consisting of a BN encoder, Flow, posterior encoder, HiFi-GAN decoder, and discriminator, as shown in Figure~\ref{fig_3}. 
The BN encoder, conditioned on the extracted emotion embedding, takes the BN feature, pitch, and energy as input to provide the prior distribution of the speaker-independent representations. The speaker embedding from the speaker lookup table with MBV is treated as the conditional constraint of the Flow model. Similar to the Text2SE module, for decoupling the speaker and emotion in the SE2Wave module, we also add classification constraints to the emotion embedding and use vCLUB to minimize mutual information between the emotion embedding and the stop gradient speaker embedding. 

%Similar to the Text2SE model, in the SE2Wave model, we utilize an emotion encoder with the MBV for information compression and the semi-supervising classification constraint to extract the emotion embedding. The PPG encoder, on the condition of emotion embedding, takes PPGs, pitch, and energy as input to provide the prior distribution of the speaker-independent representations. The speaker embedding from the speaker lookup table with the MBV is treated as the constraint of the Flow module. For decoupling the speaker and emotion in the SE2Wave model, we also use vCLUB to minimize mutual information between the emotion embedding and speaker embedding after the stop gradient.The networks of Flow, posterior encoder, HiFiGAN decoder, and discriminator follow the structure used in VITS \cite{Kim2021ConditionalVA}.

%architecture is shown in Figure \ref{fig_3}. Based on AdaVITS, we removed PPG Predictor and replaced the ISTFT decoder with the HiFiGAN \cite{Kong2020HiFiGANGA} decoder to generate high-quality audio. We add emotion information in the PPG encoder and speaker information in all affine coupling layers of the regularized flow in the prior encoder and all affine coupling layers in the HiFiGAN decoder. The emotional information is obtained from the mel-spectrogram by extracting the intermediate representation through the emotional encoder and then compressing it with MBV.

\begin{figure}[htb]
% \vspace{-8pt}

\begin{minipage}[b]{1.0\linewidth}
  \centering
  \centerline{\includegraphics[width=0.8\textwidth,height=0.8\textwidth]{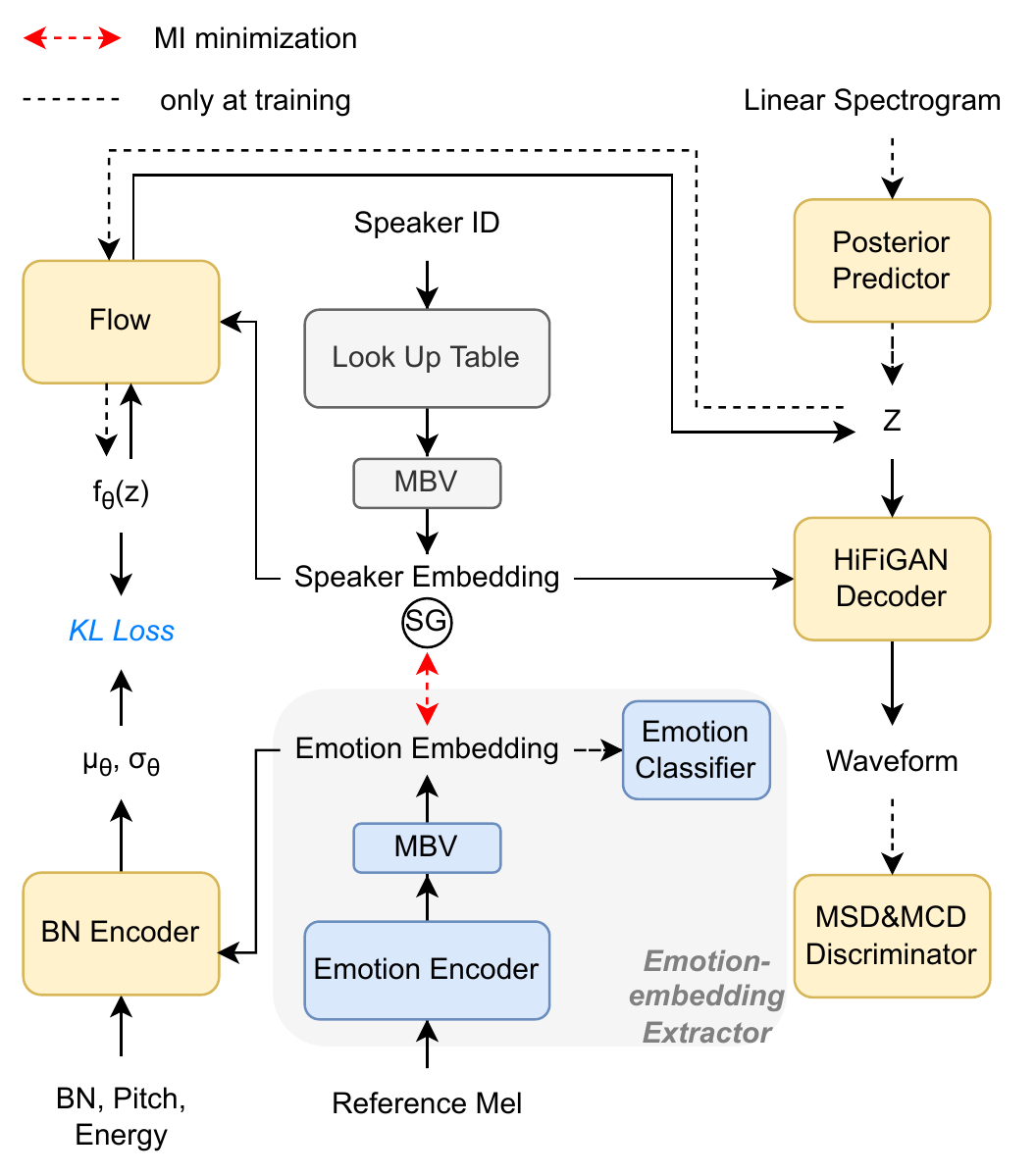}}
%  \vspace{2.0cm}
\end{minipage}
\vspace{-20pt}

\caption{The SE2Wave module architecture.}
\label{fig_3}
\vspace{-16pt}

\end{figure}

%Like the previous section's relationship between style and emotion, here we make a semi-supervised classification constraint on emotional representation and decouple emotion and speaker information based on mutual information. Because the speaker information is obtained from the Look Up Table through the speaker ID, the information is pure. While the emotional representation is obtained from the mel-spectrogram, and the information is complex. Therefore, when vCLUB evaluates mutual information and carries out gradient propagation, we make a stop gradient on the speaker representation.

We denote $\mathcal{L}_{\mathrm{emo}}$ as the emotion classification loss and $\mathcal{L}_{\mathrm{MI^{'}}}$ as the MI loss in the SE2Wave module. The L1 loss is used as the reconstruction loss $\mathcal{L}_{\mathrm{rec}}$ to minimize the mel-spectrogram of the ground truth and predicted waveform. The adversarial training loss and feature map loss in VITS~\cite{Kim2021ConditionalVA} are also adopted in the model for improving the performance of wave generation. 

The training objectives of the SE2Wave module are as follows:
{
\setlength\abovedisplayskip{2.0pt}
\setlength\belowdisplayskip{2.0pt}
\begin{equation}
\begin{aligned}  
\mathcal{L}_{\mathrm{se2w}}^{\mathrm{G}}= & \mathcal{L}_{\mathrm{kl}}+ 45 \cdot \mathcal{L}_{rec}+ 0.1 \cdot \mathcal{L}_{\mathrm{MI^{'}}} + \\
& \mathcal{L}_{\mathrm{emo}} + \mathcal{L}_{\mathrm{adv}}(\mathrm{G}) + \mathcal{L}_{\mathrm{fm}}(\mathrm{G})
\end{aligned}\label{eq:eq2}
\end{equation}
}
{
\setlength\abovedisplayskip{4.0pt}
\setlength\belowdisplayskip{2.0pt}
\begin{equation}   
\mathcal{L}_{\mathrm{se2w}}^{\mathrm{D}}=\mathcal{L}_{\mathrm{adv}}(\mathrm{D}),
\end{equation}
}
where $\mathcal{L}_{\mathrm{G}}$, $\mathcal{L}_{\mathrm{D}}$, and $\mathcal{L}_{\mathrm{kl}}$ are the generative loss, discriminator loss, and KL divergence of the hidden distribution.

\vspace{-10pt}
\subsection{Semi-supervised training}
\vspace{-5pt}

%Suppose we have a variety of multi-speaker expressive data at hand, including emotion-labelled data, style-labelled data and unlabeled data. To better make use of the data, we introduce a semi-supervised training strategy to train the emotion/style embedding extractors for multi-label classification. Specifically, for the emotion-labelled data, only emotion classification loss is calculated, and the model softly determines what style it should be. Likewise for the style-labelled data.
Suppose we have a variety of multi-speaker expressive data at hand, including emotion-labeled data, style-labeled data, and unlabeled data. To better leverage all of the data, we introduce a simple semi-supervised training strategy to train the emotion/style embedding extractors for multi-label classification. Specifically, for the emotion-labeled data, only $\mathcal{L}_{\mathrm{emo}}$ is calculated and $\mathcal{L}_{\mathrm{sty}}$ is not considered in Eq.~(\ref{eq:eq1}); the model softly determines what style it should be. Likewise for the style-labeled data. As for the unlabeled data, neither $\mathcal{L}_{\mathrm{emo}}$ nor $\mathcal{L}_{\mathrm{sty}}$ is calculated in Eq.~(\ref{eq:eq1}) and Eq.~(\ref{eq:eq2}), and the model will softly determine what style or emotion it contains.

\vspace{-10pt}
\subsection{Attention based reference selection}
\vspace{-5pt}

At training time, the style/emotion embedding is extracted from the target mel-spectrogram, which is parallel with the text content. While during inference, the reference signal is different from the novel text (i.e. non-parallel). This results in a mismatch problem of the extracted embedding between the training and inference stages, leading to degraded performance according to previous studies~\cite{Bian2019MultireferenceTB}.
To relieve the performance degradation brought by the mismatch, we introduce an extra stage to fine-tune the Text2SE and SE2Wave modules. Particularly in the fine-tuning stage, we introduce a novel embedding extractor to replace the original emotion/style extractor used in the previous training stage. The new extractor aims to alleviate the aforementioned mismatch problem and select the appropriate reference in a soft way. Thus the new embeddings are used as conditions for the Text2SE and SE2Wave modules.

As illustrated in Figure~\ref{fig_3}, the embedding extractor in fine-tuning procedure employs scaled dot-product attention~\cite{DBLP:conf/nips/VaswaniSPUJGKP17} to calculate the embedding output as the conditional constraints for the Text2SE and SE2Wave modules. We introduce an \textit{embedding-candidate pool} providing the candidates as the attention \textit{key} and \textit{value}, retrieved from the given style/emotion ID. Specifically, the embedding-candidate pool consists of $N$ embeddings for each category (e.g., emotion-happy or style-fairy-tales) extracted by the style/emotion encoder and MBV in the previous training stage. Unlabeled data is treated as a special category. The attention \textit{query} is provided by a GST-layer~\cite{DBLP:conf/icml/WangSZRBSXJRS18} from the input hidden representations which are the phoneme encoder outputs in the Text2SE module and BN, pitch, and energy inputs in the SE2Wave module respectively. 
In this way, the attention mechanism selects the embedding with the most
significant attention weight as  the appropriate reference embedding for the models based on the linguistic input during fine-tuning.

\begin{figure}[htb]
\vspace{-8pt}

\begin{minipage}[b]{1.0\linewidth}
  \centering
  \centerline{\includegraphics[width=0.78\textwidth]{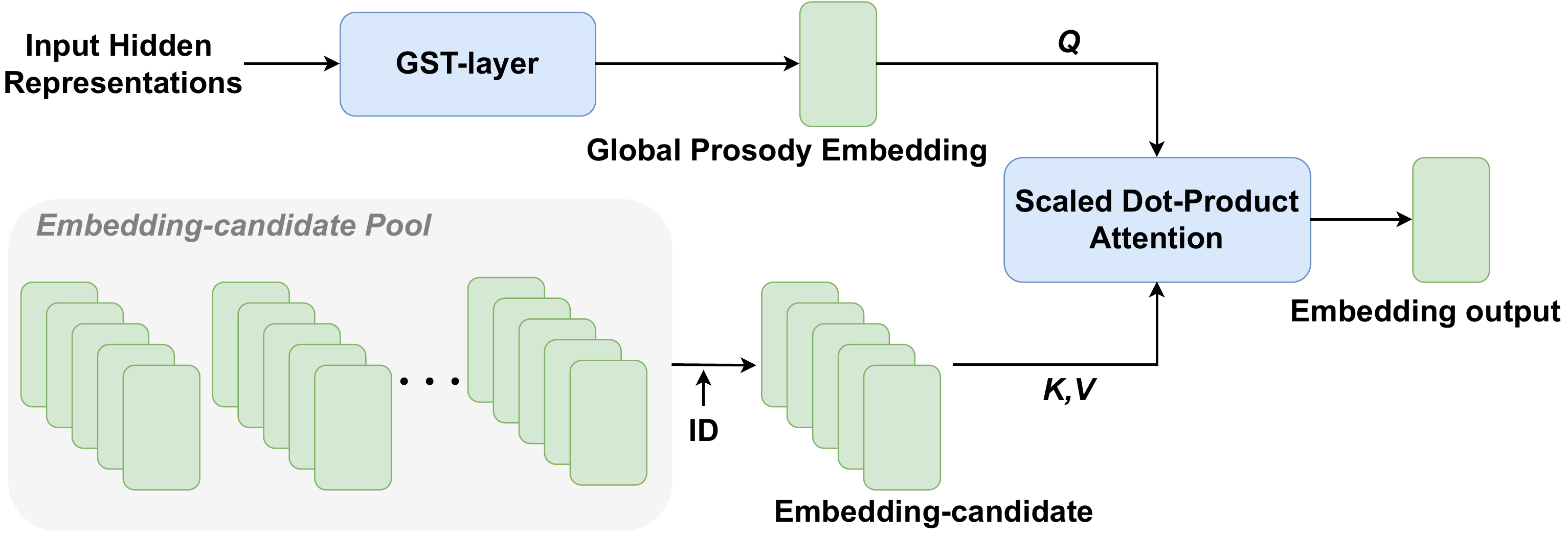}}
%  \vspace{2.0cm} 
\end{minipage}
\vspace{-16pt}

\caption{The embedding extractor in fine-tuning procedure.}
\label{fig_3}
\end{figure}
\vspace{-8pt}

\vspace{-10pt}
\subsection{Pipeline}
\vspace{-5pt}

The pipeline of the proposed approach contains training, fine-tuning, and inference phases. During training, we train the two-stage model using the objectives mentioned above and update the emotion-/style-embedding extractors $E_{emo}^{t}$ and $E_{sty}^{t}$ by the extracted embeddings with the constraint of MI loss and the semi-surprised classification loss. During fine-tuning, the embedding extractors $E_{emo}^{f}$ and $E_{sty}^{f}$ are updated by the objectives of the acoustic model. In inference, the proposed two-stage system adopts the same embedding extractor as the fine-tuning stage. Both the Text2SE and SE2Wave modules automatically select the appropriate embedding according to the textual input, style, and emotion.

%In inference, the proposed two-stage system adopts the embedding extractor as the same as fine-tuning time. The system takes the text, style id, emotion id, and speaker id as input to generate the speaker-independent BN by the Text2SE model and waveform in desired multiple factors (i.e. speaker, style, and emotion) by the SE2Wave model. Both Text2SE and SE2Wave models automatically select the appropriate embedding according to the textual content, style, and emotion.

\vspace{-10pt}
\section{EXPERIMENTS}
\vspace{-10pt}
\label{sec:typestyle}

\begin{table*}[]
\centering
\caption{Results of subjective evaluation  with 95$\%$ confidence interval and objective evaluation.}
\footnotesize
    \label{tab:exp}
\begin{tabular}{c|cccc|cc}
\toprule
Model    & Naturalness & Emotion Similarity & Speaker Similarity & Style Similarity & CER~($\%$) & Cosine Similarity \\ \midrule
Proposed & \textbf{4.03 $\pm$ 0.08}            & \textbf{3.89 $\pm$ 0.10}                   & \textbf{3.93 $\pm$ 0.07} & \textbf{3.81 $\pm$ 0.11}                  & 6.7    & \textbf{0.856}             \\
MR-Tacotron~\cite{Bian2019MultireferenceTB}      & 3.83 $\pm$ 0.09            & 3.38 $\pm$ 0.12                   & 3.62 $\pm$ 0.10   & 3.77 $\pm$ 0.10                & \textbf{6.0}    & 0.804             \\
Referee~\cite{Liu2022RefereeTR}      & 3.07 $\pm$ 0.11            & 2.88 $\pm$ 0.12                   & 3.37 $\pm$ 0.08 & 3.43 $\pm$ 0.11                  & 7.6    & 0.846             \\ \bottomrule
\end{tabular}
\vspace{-16pt}
\end{table*}

\subsection{Experimental Setup}
\vspace{-5pt}

Three internal Mandarin corpora are involved in the experiments: 1) dataset \textbf{M30S3} contains 30 speakers, and its total duration is 18.5 hours, in which each speaker has 1 to 3 styles including \textit{poetry recitation}, \textit{story telling - fairy tales} and \textit{story telling - novels}; 2) dataset \textbf{M3E6} contains three speakers, and its total duration is 21.1 hours, in which each speaker has six emotions of \textit{anger, fear, happiness, sadness, surprise and neutral}; and 3) dataset \textbf{M30U} has 30 speakers with neither style tags nor emotion tags, and its total duration is 18.2 hours. For all recordings, the sample rate is converted to 24 kHz. The BN features are extracted with 12.5ms hop-size and 256-dimension from a robust TDNN-F model trained with 30k hours of Mandarin speech data. To validate the performance of our proposed approach, we implement the following systems:
\begin{itemize}
\vspace{-4pt}
    \item \textbf{MR-Tacotron}: Multi-reference structure follows~\cite{Bian2019MultireferenceTB} to disentangle and control specific styles based on the FastSpeech~2 architecture for a fair comparison. A  HiFi-GAN~\cite{DBLP:conf/nips/KongKB20} vocoder is adopted to transform the predicted mel-spectrogram into speech waveform.
    \vspace{-4pt}
    \item \textbf{Referee}: A cross-speaker style transfer framework follows Referee~\cite{Liu2022RefereeTR} with additional emotion transfer.
    \vspace{-4pt}
    \item \textbf{Proposed}: the two-stage framework proposed in this paper to decouple and recompose the multiple factors in speech.
\vspace{-4pt}

\end{itemize}
In our implementation of the proposed approach, the emotion and style encoders follow the structure of mel-style encoder proposed by Meta-StyleSpeech~\cite{Min2021MetaStyleSpeechM}. All classifiers have the same structure that consists of 3 fully connected layers. The Text2SE backbone and mutual information estimator follow the settings of FastSpeech~2~\cite{Ren2021FastSpeech2F}, and VQMIVC~\cite{Wang2021VQMIVCVQ} respectively. The SE2Wave backbone follows the settings of VITS~\cite{Kim2021ConditionalVA}, and the BN encoder consists of 6 FFT blocks. During fine-tuning, we set $N=100$ for the embedding-candidate pool in this paper.

% \usepackage{booktabs}
% \begin{table}[]
% \centering
% \caption{Results of evaluations on speech naturalness, emotion similarity, and speaker similarity. ``SN", ``ES", ``SS", and ``SCS" denote speech naturalness, emotion similarity, speaker similarity, and speaker cosine similarity.}
%     \label{fig_3-2-1}
% \begin{tabular}{c|ccc|c}
% \toprule
% model    & SN & ES & SS & SCS\\ \midrule
% Proposed & \textbf{4.03 $\pm$ 0.08}            & \textbf{3.89 $\pm$ 0.10}                   & \textbf{3.93 $\pm$ 0.07}   & \textbf{0.856} \\
% MR-Tacotron      & 3.83 $\pm$ 0.09            & 3.38 $\pm$ 0.12     & 3.62 $\pm$ 0.10  & 0.804     \\
% Referee      & 3.07 $\pm$ 0.11            & 2.88 $\pm$ 0.12   & 3.37 $\pm$ 0.08  & 0.846      \\ \bottomrule
% \end{tabular}
% \vspace{-10pt}
% \end{table}

\vspace{-10pt}
\subsection{Subjective evaluation}
\vspace{-5pt}

We conduct mean opinion score (MOS) experiments to evaluate speech naturalness, emotion similarity, speaker similarity, and style similarity. Specifically, given 20 reserved transcripts for each style, we generate samples respectively for each emotion category, resulting in 360 listening samples ($20\times3 \times6$). We invite 20 native Mandarin Chinese speakers to participate in the listening tests. In each test session (naturalness/emotion/speaker/style), participants are asked to rate how similar the synthetic and the reference speech is in the specific assessment metric while ignoring other aspects. 

As shown in Table~\ref{tab:exp}, the proposed approach significantly outperforms Referee and MR-Tacotron in terms of speech naturalness, emotion similarity, and speaker similarity. The high emotion and speaker similarity scores demonstrate that the proposed approach can decouple the emotion and speaker identity effectively. For the style similarity, the proposed approach achieves much better performance than Referee and is slightly better than MR-Tacotron. The results of the emotion, style and speaker similarity indicate that the proposed method can effectively decouple multiple factors from speech.
Besides, the proposed method achieves the best audio naturalness, indicating its flexibility and stability in generating specific emotional speech of target speakers.

\vspace{-10pt}
\subsection{Objective evaluation}
\vspace{-5pt}

%The objective experiments are conducted to evaluate the performance of models and the extracted embeddings are visualised for intuitive observations.

%\begin{table}[]
%\centering
%\caption{Results of objective evaluation}
%\label{tab:obj}
%\begin{tabular}{@{}c|ccc@{}}
%\toprule
%model  & CER & Cosine Similarity \\ \midrule
%Proposed   & 6.70 $\%$  & \textbf{0.856}  \\
%MR-Tacotron   & \textbf{5.96 $\%$}  & 0.804 \\
%Referee    & 7.56 $\%$   & 0.846  \\ \bottomrule
%\end{tabular}
%\vspace{-10pt}
%\end{table}

% \textbf{Speaker Cosine Similarity}. 
% We use the pre-trained ECAPA-TDNN~\cite{Desplanques2020ECAPATDNNEC} to extract the x-vector and calculate the cosine similarity to evaluate the speaker similarity, as shown in Table~\ref{fig_3-2-1}.
\textbf{Robustness}.
We measure the character error rate (CER) of the synthesized samples by the pre-trained WeNet~\cite{Yao2021WeNetPO} to assess the robustness of the models. Moreover, we use the pre-trained ECAPA-TDNN~\cite{Desplanques2020ECAPATDNNEC} to extract the x-vector and calculate the cosine similarity to further verify the speaker similarity. As shown in Table~\ref{tab:exp}, the proposed model achieves the highest cosine similarity. Interestingly, the Referee achieves a similar cosine similarity to the proposed method. However, subjective tests show a massive gap between the proposed method and the Referee in speaker similarity. We speculate that the poor audio quality of the Referee affects the listeners' judgment. We observe that the ASR model does not recognize Poet well, where CER is generally high. Considering speech generated from the proposed method is expressive of the multiple factors, it's reasonable to get a slightly higher CER than MR-Tacotron.

\begin{figure}[htb]
\vspace{-10pt}

\begin{minipage}[b]{0.327\linewidth}
  \centering
  \centerline{\includegraphics[width=\textwidth]{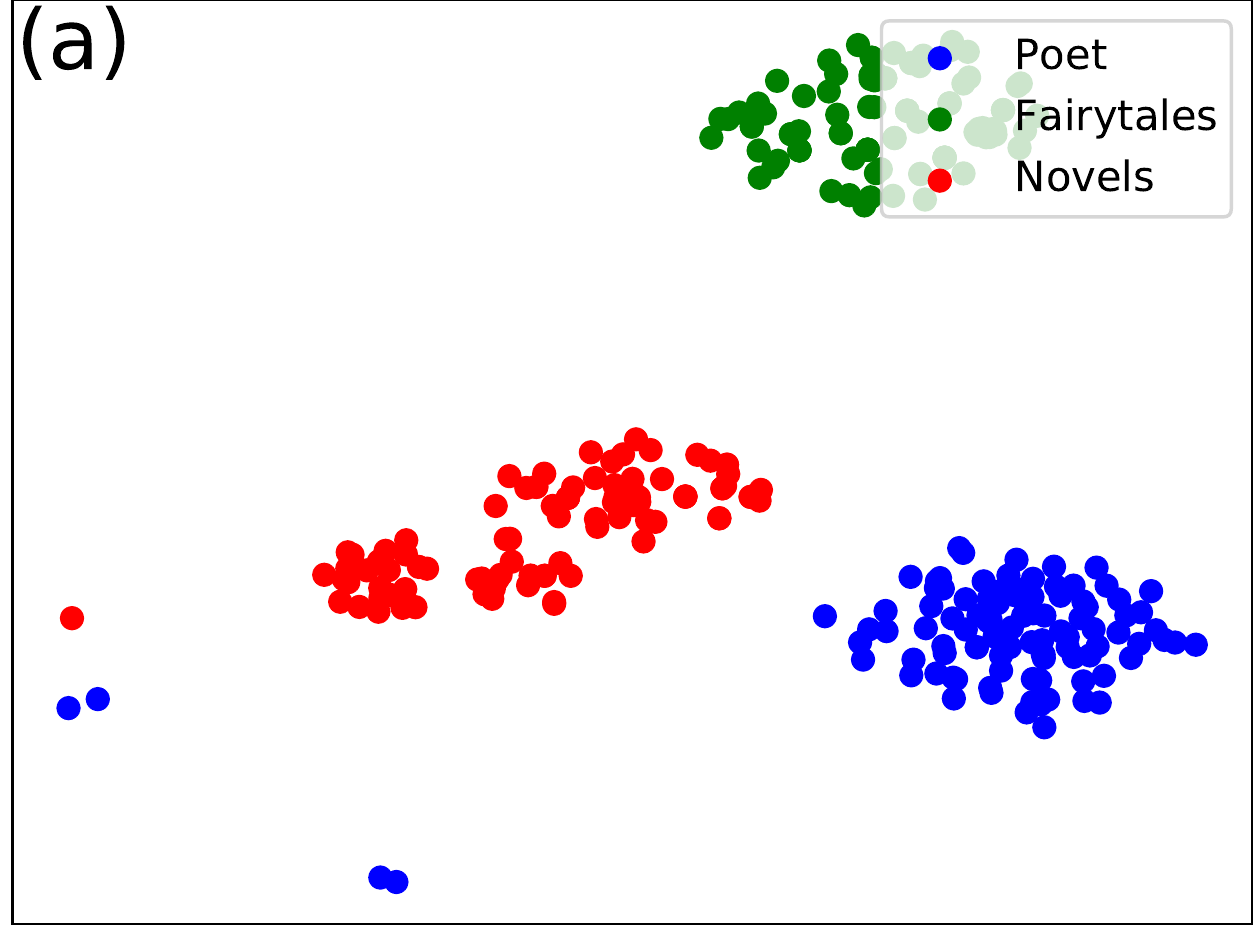}}
%   \centerline{(a)}\medskip
\end{minipage}
\hfill
\begin{minipage}[b]{0.327\linewidth}
  \centering
  \centerline{\includegraphics[width=\textwidth]{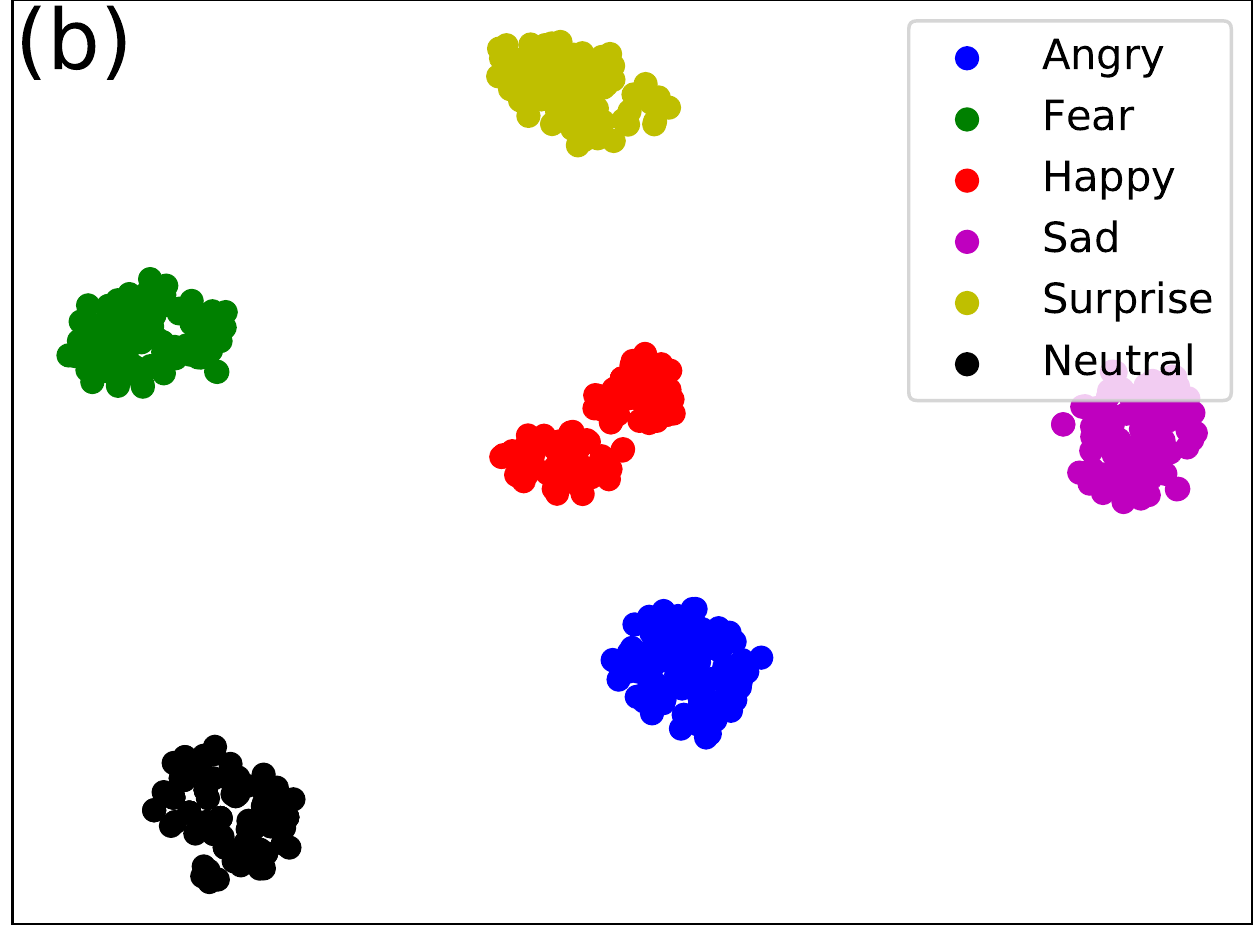}}
%  \vspace{1.5cm}
%   \centerline{(b)}\medskip
\end{minipage}
\hfill
\begin{minipage}[b]{0.327\linewidth}
  \centering
  \centerline{\includegraphics[width=\textwidth]{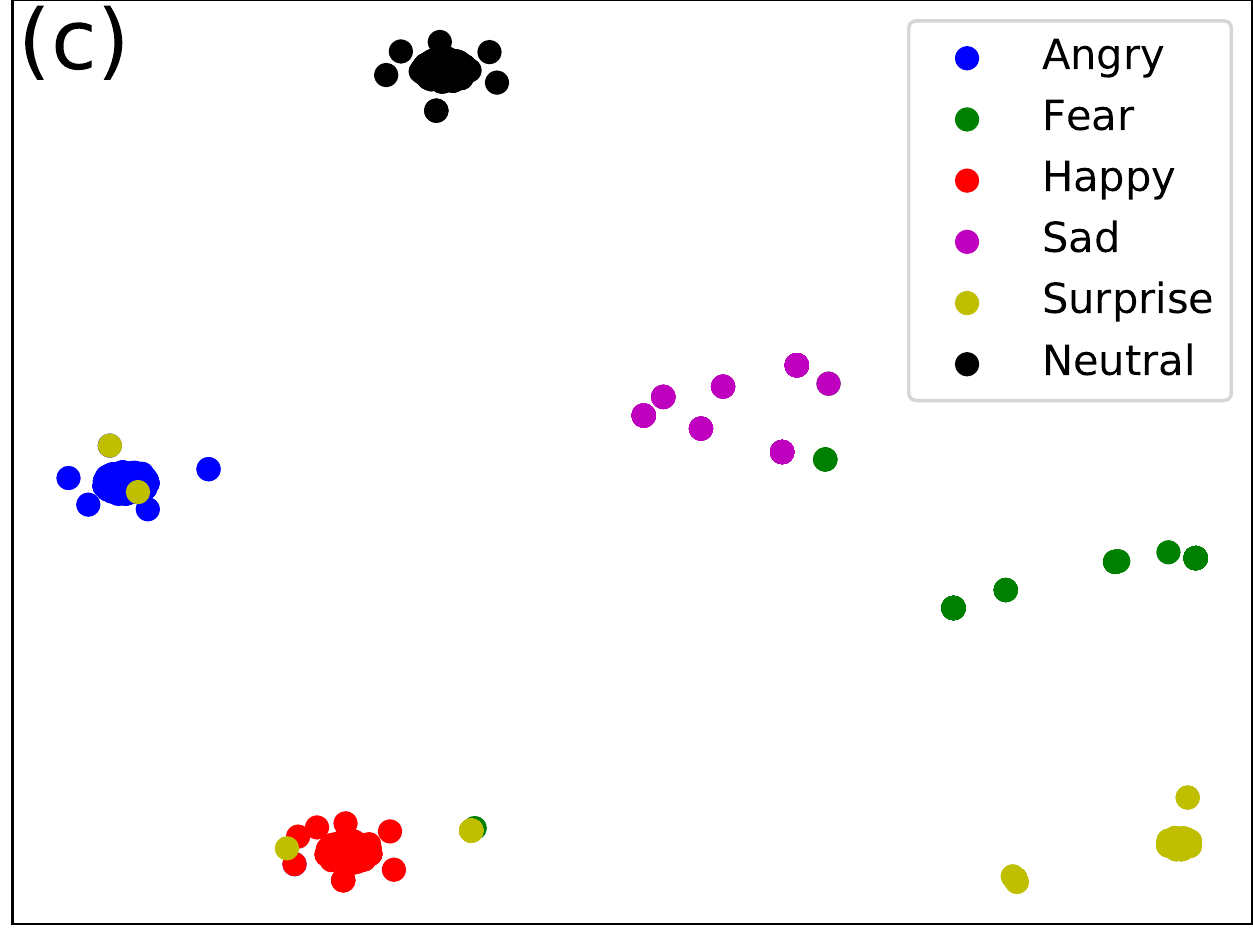}}
%  \vspace{1.5cm}
%   \centerline{(c)}\medskip
\end{minipage}
\vspace{-20pt}

\caption{T-SNE results of style embedding and emotion embedding. (a) style embedding in Text2SE, (b) emotion embedding in Text2SE and (c) emotion embedding in SE2Wave.}
\label{fig_3-3-2}
\vspace{-8pt}
\end{figure}

\textbf{Effectiveness of semi-supervised training}. To verify the effectiveness of semi-supervised training, we visualize the emotion and style embeddings through t-SNE~\cite{Maaten2008VisualizingDU}. One hundred utterances reserved per emotion or style are adopted for the test. As shown in Figure~\ref{fig_3-3-2}(a) and  \ref{fig_3-3-2}(b), the style and emotion embeddings are well clustered, proving the effectiveness of semi-supervised training. Moreover, Figure~\ref{fig_3-3-2}(c) demonstrates that emotion embeddings in the SE2Wave model cannot form clusters by categories. We conjecture that emotion embeddings in SE2Wave mainly focus on and supplement the fine-grained emotional variance since emotion in Text2SE represents the global category.

\begin{figure}[htb]
\vspace{-5pt}
\begin{minipage}[b]{0.05\linewidth}
    \leftline{(a)}
\end{minipage}
\hfill
\begin{minipage}[b]{0.9\linewidth}
  \leftline{\includegraphics[width=\textwidth]{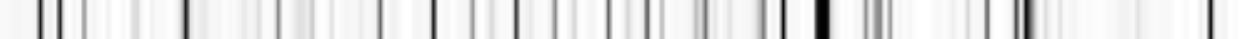}}
\end{minipage}

\begin{minipage}[b]{0.05\linewidth}
    \leftline{(b)}
\end{minipage}
\hfill
\begin{minipage}[b]{0.9\linewidth}
  \leftline{\includegraphics[width=\textwidth]{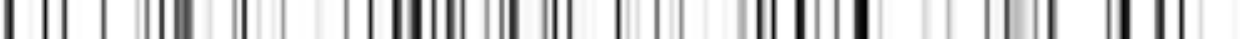}}
\end{minipage}

\begin{minipage}[b]{0.05\linewidth}
    \leftline{(c)}
\end{minipage}
\hfill
\begin{minipage}[b]{0.9\linewidth}
  \leftline{\includegraphics[width=0.75\textwidth]{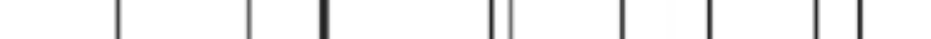}}
\end{minipage}
\vspace{-10pt}

\caption{Results of the standard deviation of each dimension of the extracted embedding. (a) style embedding in Text2SE, (b) emotion embedding in Text2SE and (c) emotion embedding in SE2Wave.}
\label{fig_3-3-3}
\vspace{-8pt}

\end{figure}

\textbf{Emotion and Style Embedding}. We extract embeddings of all training utterances and then calculate the standard deviation of all embeddings for each dimension. The lower standard deviation means the dimension changes less among embeddings of all utterances, which is less valuable, in other words. The embeddings are intuitively presented in Figure~\ref{fig_3-3-3}, where the darker vertical bar means a higher standard deviation. We can observe that some dimensions are purely white, meaning they are not used at all. Moreover, Figure~\ref{fig_3-3-3}(c) shows obviously fewer dimensions are used in the SE2Wave emotion embedding since it is just a supplement to provide fine-grained emotional variance. These observations show that MBV can effectively discretize information in the embeddings. More importantly, Figure~\ref{fig_3-3-3}(a) and (b) are mutually exclusive, indicating that style and emotion are well decoupled.
%We calculate the standard deviation of each dimension of the style embedding and emotion embedding of all datasets' utterances and then maximum normalization to normalize results to [0,1]. Figure~\ref{fig_3-3-3} shows the results, the darker the colour, the greater the calculated value. Some results are pure white, meaning these dimensions never changed. In other words, they are not used. Moreover, Figure~\ref{fig_3-3-3}(c) shows fewer dimensions are used in SE2Wave emotion embedding. The above results show that MBV can prevent redundant content from leaking and not over-compress.

\begin{table}[]
\vspace{-6pt}
\setlength\tabcolsep{5pt}
\footnotesize
\centering
\caption{Results of Ablation study with 95$\%$ confidence interval.}
    \label{fig_3-4-1}
\begin{tabular}{l|cccc}
\toprule
Model       & Naturalness & \begin{tabular}[c]{@{}c@{}}Emotion \\ Similarity\end{tabular} & \begin{tabular}[c]{@{}c@{}}Speaker \\ Similarity\end{tabular} & \begin{tabular}[c]{@{}c@{}}Style\\ Similarity\end{tabular} \\ \midrule
Proposed    & \textbf{4.03 $\pm$ 0.08}            & \textbf{3.89 $\pm$ 0.10}                   & \textbf{3.93 $\pm$ 0.07} & \textbf{3.81 $\pm$ 0.11}     \\
- MBV     & 3.47 $\pm$ 0.10                   & 3.68 $\pm$ 0.08                          & 3.33 $\pm$ 0.10 & 3.60 $\pm$ 0.14   \\
- MI      & 3.74 $\pm$ 0.10                   & 3.88 $\pm$ 0.08                          & 3.21 $\pm$ 0.11 & 3.71 $\pm$ 0.14    \\
- FT      & 3.88 $\pm$ 0.08                   & 3.83 $\pm$ 0.09                          & 3.63 $\pm$ 0.09 & 3.76 $\pm$ 0.13    \\
- EE      & 3.85 $\pm$ 0.08                   & 3.49 $\pm$ 0.09                          & 3.90 $\pm$ 0.08 & 3.78 $\pm$ 0.14    \\ 
- BN      & 2.63 $\pm$ 0.12                   & 2.47 $\pm$ 0.13                          & 2.20 $\pm$ 0.11 & 2.38 $\pm$ 0.15    \\ \bottomrule
\end{tabular}
\vspace{-16pt}
\end{table}

\vspace{-12pt}
\subsection{Ablation study}
\vspace{-5pt}
We conduct an ablation study to validate the components of our proposed method by removing certain modules and jointly training two modules, as shown in Table~\ref{fig_3-4-1}. The results show that removing the Multi-label Binary Vector (-MBV) module leads to a decline in perceptive evaluations, indicating that MBV improves system stability. Removing the Mutual information loss (-MI) and the emotion extractor in SE2Wave (-EE) lead to a sharp decline in speaker similarity and emotion similarity, respectively, highlighting the effectiveness of MI loss in decoupling multiple factors and the importance of fine-grained emotional variance composition in the SE2Wave stage. The results (-FT) also show that the fine-tuning process can effectively improve overall performance. Furthermore, the jointly trained model (-BN) fails to disentangle multiple factors, resulting in synthetic audio with low naturalness, expressiveness, and speaker similarity.

%We further conduct an ablation study to validate the components in the proposed method by removing Multi-label Binary Vector (MBV), Mutual information (MI), fine-tuning (FT), emotion extractor in SE2Wave (EE) and BN features and jointly train two modules (BN), as shown in Table~\ref{fig_3-4-1}. Without the MBV module, the system performs an overall decline in perceptive evaluations, which verifies the MBV module improves the system stability. Removing the MI loss and emotion extractor in SE2Wave respectively results in a sharp decline in speaker similarity and emotion similarity. These results prove the effectiveness of MI loss to decouple the multiple factors and the necessity of fine-grained emotional variance composition in the SE2Wave stage. In addition, the results of "-FT" demonstrate that fine-tuning process can effectively improve overall performance. Furthermore, the jointly trained model fails disentangle multiple factors, where the synthetic audio has low naturalness, expressiveness and speaker similarity.
%the model becomes unstable, resulting in an overall decline in performance. Without the MI module, multiple factors are coupled, resulting in a sharp decline in speaker similarity. Without the EE module, the audio's emotional similarity decreases significantly, which proves the necessity of replenishing emotional details in SE2Wave. After fine-tuning, the mismatch between train and inference is eliminated, which leads to better performance. The above results demonstrate the necessity of the modules used in the proposed method.

\vspace{-10pt}
\section{Conclusions}
\vspace{-8pt}
\label{sec:majhead}
This paper aims to synthesize speech with desired style and emotion for a target speaker by transferring the style and emotion from reference speech recorded by other speakers.
We approach this challenging problem with a two-stage framework composed of a text-to-style-and-emotion (Text2SE) module and a style-and-emotion-to-wave (SE2Wave) module, while neural bottleneck features are served as the intermediate representation. Importantly, based on this framework, we have proposed  several contributions, including multi-factor decomposition, semi-supervised training to better leverage data, and attention-based reference selection. Extensive experiments demonstrate the good design of our model. %Future work includes extending the proposed work for synthesizing highly expressive speech with a specific style and emotion for unseen target speakers.

\vfill\pagebreak
\clearpage
% \section{REFERENCES}
% \label{sec:refs}

% List and number all bibliographical references at the end of the
% paper. The references can be numbered in alphabetic order or in
% order of appearance in the document. When referring to them in
% the text, type the corresponding reference number in square
% brackets as shown at the end of this sentence \cite{C2}. An
% additional final page (the fifth page, in most cases) is
% allowed, but must contain only references to the prior
% literature.

% \bibliographystyle{IEEEtran}

\bibliographystyle{IEEE}
\bibliography{strings,refs}

\end{document}